\documentclass[twocolumn,showpacs,preprintnumbers,
amsmath,amssymb,aps,prd,nofootinbib,superscriptaddress,
eqsecnum]{revtex4}

\usepackage{graphicx,color}

\makeatletter
\renewcommand{\p@subsection}{}
\makeatother

\begin{document}

\title{
 Net-proton probability distribution in heavy ion  collisions
 }

\author{P. Braun-Munzinger}
\affiliation{%
ExtreMe Matter Institute EMMI, GSI, D-64291 Darmstadt, Germany}
\affiliation{%
GSI Helmholtzzentrum f\"ur Schwerionenforschung, D-64291
Darmstadt, Germany }
\affiliation{Technical
University, D-64289 Darmstadt,  Germany}
\affiliation{Frankfurt Institute for Advanced Studies, J.W.
Goethe University, Frankfurt, Germany}

\author{B. Friman}
\affiliation{%
GSI Helmholtzzentrum f\"ur Schwerionenforschung, D-64291
Darmstadt, Germany }
\author{F. Karsch}
\affiliation{%
Physics Department, Brookhaven National Laboratory, Upton, NY 11973, USA
}
\affiliation{%
Fakult\"at f\"ur Physik, Universit\"at Bielefeld,
 D-33501 Bielefeld, Germany
}
\author{K. Redlich}
\affiliation{%
ExtreMe Matter Institute EMMI, GSI, D-64291 Darmstadt, Germany}
\affiliation{%
Institute of Theoretical Physics, University of Wroclaw, PL--50204
Wroc\l aw, Poland}
\author{V. Skokov}
\affiliation{%
GSI Helmholtzzentrum f\"ur Schwerionenforschung, D-64291
Darmstadt, Germany }
\affiliation{Frankfurt Institute for Advanced Studies, J.W.
Goethe University, Frankfurt, Germany}

\date{\today}

\begin{abstract}
We compute  net-proton probability distributions in heavy ion collisions
within the hadron resonance gas model.  The model results are compared with
data taken by the STAR Collaboration in Au-Au collisions at
$\sqrt{ s_{NN}}=$ 200 GeV for different centralities. We show that in
peripheral  Au-Au collisions  the measured distributions, and the resulting
first four moments of net-proton fluctuations, are consistent with results
obtained from the hadron resonance gas model.
However, data taken in central Au-Au collisions differ from the predictions of the model.
The observed deviations can not be attributed to uncertainties in
model parameters.
We discuss possible interpretations of the observed
deviations.

\end{abstract}


\setcounter{footnote}{0}

\maketitle


\section{Introduction}

One of the  objectives  of heavy ion experiments at CERN and BNL
is to probe
 properties of
the  QCD phase diagram related to deconfinement and chiral symmetry
restoration. To experimentally verify the phase change in a
medium created in such collisions, one needs observables that are sensitive
to critical behavior. Modifications in the magnitude of fluctuations or in the
corresponding susceptibilities of conserved charges have been suggested as
possible signals for chiral symmetry restoration and deconfinement 
\cite{st,fk1,st1,karsch2,karsch31,spinodal}.

Fluctuations of baryon number and electric charge diverge at the
hypothetical critical end point in the QCD phase diagram at non-zero
temperature and baryon chemical potential, while they remain
finite along the cross-over boundary.
Consequently, large fluctuations of baryon number and electric charge as well
as a non monotonic behavior of these fluctuations as function of
the collision energy in heavy ion collisions have been proposed as a signature for the QCD critical end point
\cite{st,st1,spinodal}.

It also has been argued that even in the absence of a critical end point,
fluctuations of conserved charges, and
their higher order cumulants,
can be used to identify the phase boundary. It is expected that the fluctuations are modified if the chemical freeze-out, i.e.,
the generation of hadrons and their fluctuations, occurs shortly after
the system passed through a region where quarks were
deconfined and chiral symmetry was partially
restored \cite{karsch31,karsch3,new}.
Thus, fluctuations in the hadronic phase can show structures attributed 
to critical dynamics at the restoration of chiral
symmetry.

The critical region characterizing the cross-over transition in
the QCD phase diagram is expected to be located close
to the freeze-out curve extracted from heavy ion experiments \cite{wetterich}.
On this phenomenologically determined curve all particle yields
achieve their measured  values \cite{hwa,anton,cleymans,star3}.
Thermodynamics at freeze-out
 is, to a first approximation, well described
by the hadron resonance gas (HRG) model,
which was shown to be very successful in describing not only 
the data on particle yields
but also
the thermodynamics of a strongly interacting medium
at low temperature as computed in lattice QCD
\cite{fk1,tawfik,fk2,ratti,peter,muller}.


\begin{figure*}
\includegraphics*[width=5.5cm]{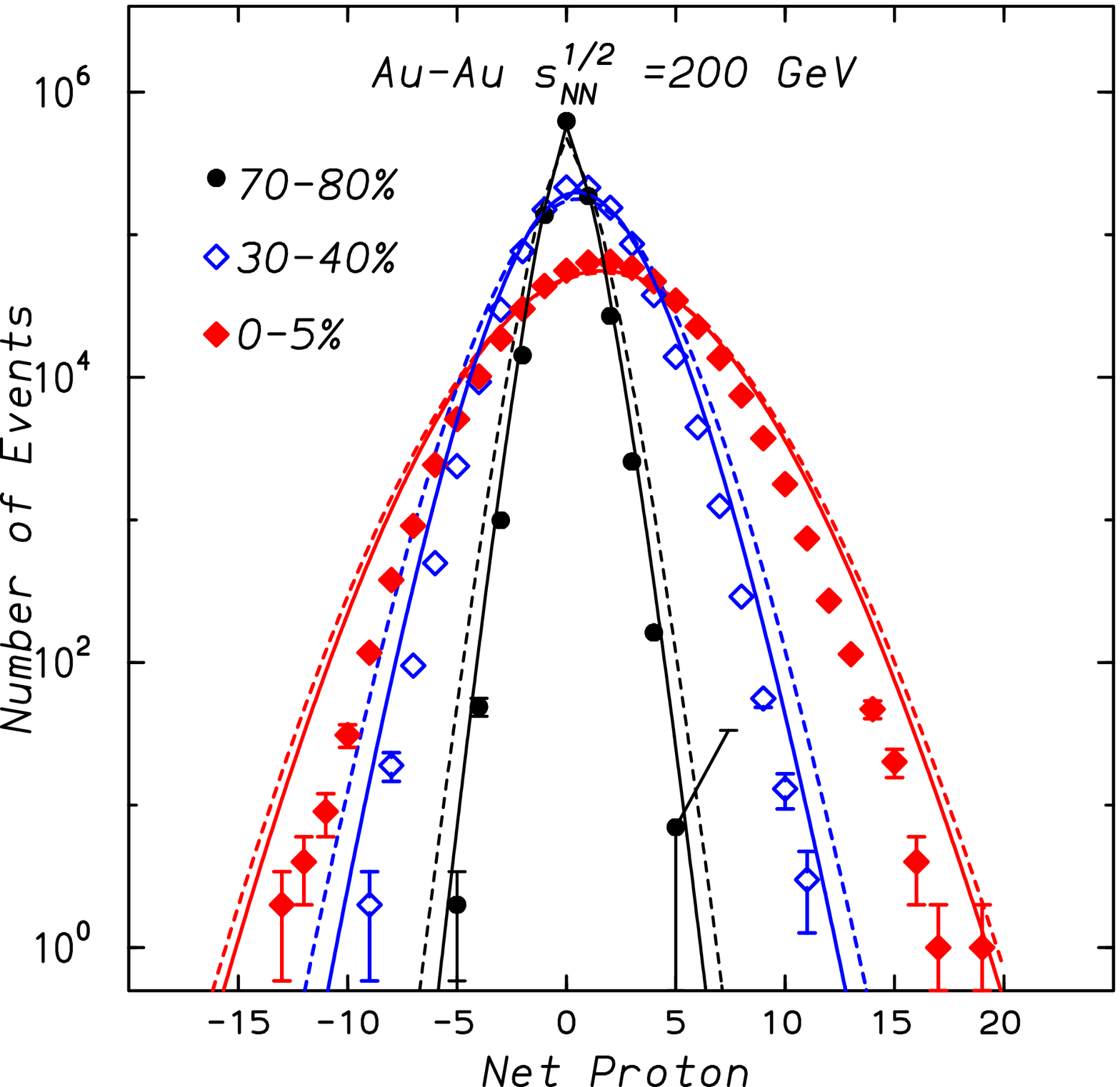}
\hspace*{1.5cm}
\includegraphics*[width=5.33cm]{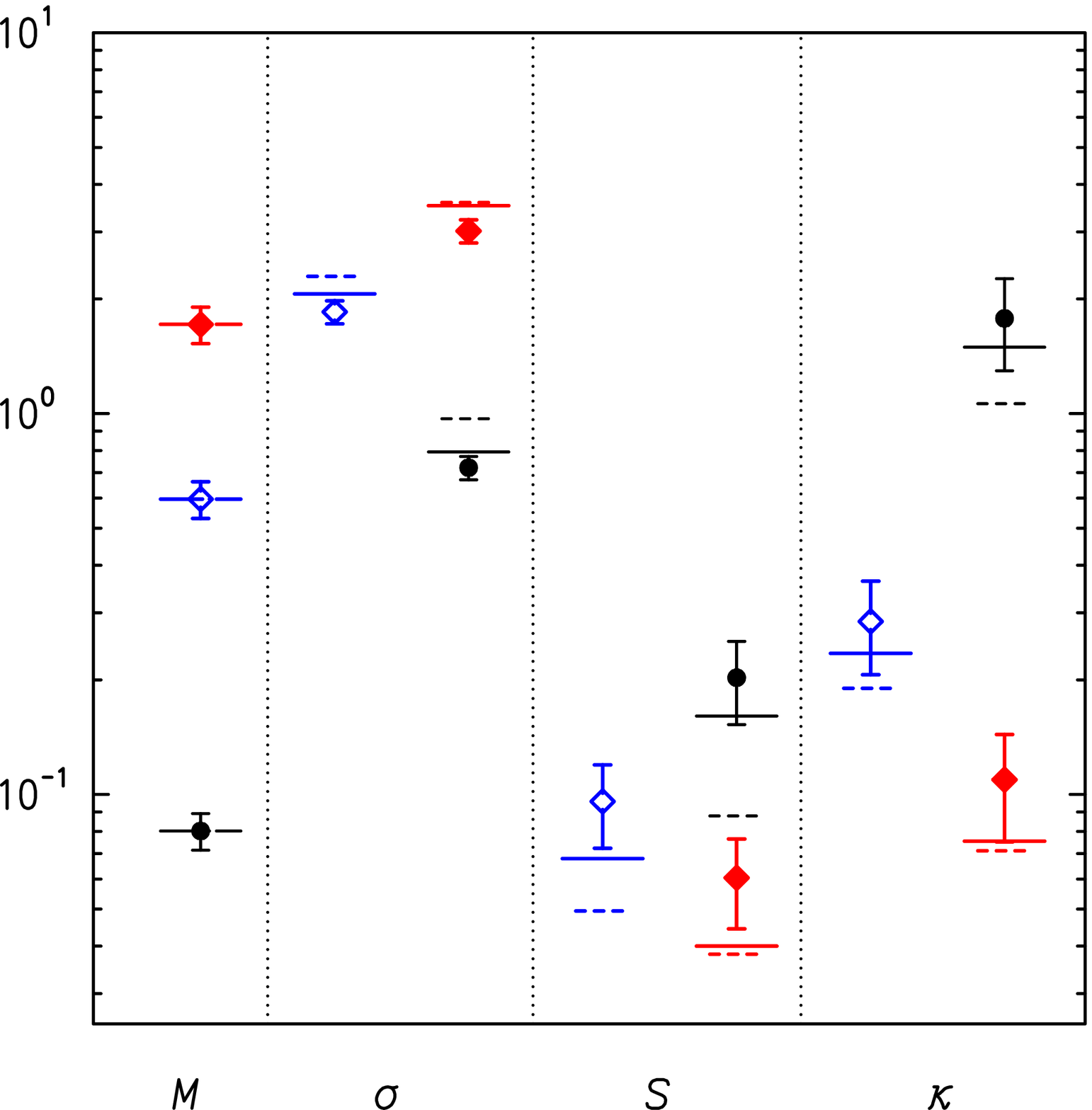}
\caption
{
Left-hand figure: Net-proton   distributions calculated in the hadron resonance gas using Eq.~\ref{eq14} and compared with STAR data for Au-Au collisions at $\sqrt s_{NN}=$ 200 GeV for different   centralities \cite{star1}.
The full-lines are obtained with experimental inputs for proton
$\langle N_p\rangle $ and anti-protons $\langle N_{\bar p}\rangle$
 yields, while the broken-lines are obtained with $\langle N_p\rangle $ and  $\langle N_{\bar p}\rangle$  computed in the thermal model with the parameters at chemical freeze-out  taken from  Ref. \cite{star3}. Right-hand figure:
Mean (M), variance ($\sigma$), skewness (S) and kurtosis ($\kappa$)
calculated from the probability distributions shown in Fig.~\ref{fig1} (left) for different centralities. Data are from  STAR  \cite{star1}.}
\label{fig1}
\end{figure*}

If chemical freeze-out occurs near or at the QCD phase boundary,
this should  be reflected in the higher order cumulants of charge fluctuations
since the sensitivity to critical dynamics grows with increasing order~\cite{karsch3}.
Consequently, the values of higher order cumulants can differ significantly
from the results of the HRG along the freeze-out curve
even if lower order cumulants agree.
In particular,  at vanishing chemical potential, the sixth and higher
order cumulants can even be negative
in the hadronic phase while the HRG yields positive values everywhere
\cite{karsch3}. In fact, the results of the HRG model
on cumulants of charge fluctuations can serve as a theoretical baseline
for the analysis of heavy ion collisions \cite{fk1,tawfik,karsch3}.
In equilibrium, any deviation from the HRG model would be a
reflection of genuine QCD properties not accounted for by the model and
could constitute evidence for critical phenomena at the time of hadronization.

Recently first  data on charge fluctuations and higher order
cumulants, identified through
net-proton fluctuations, were obtained by the STAR Collaboration
in Au-Au collisions at several collision energies \cite{star1,star2}.
To explore possible signs of
criticality, the STAR data on the first four cumulants were compared to
HRG \cite{karsch3,star2} and lattice QCD \cite{GG,schmidt} results. The basic
properties of the measured fluctuations and ratios of cumulants are
consistent with the expectations based on HRG as well as on lattice QCD
calculations.
However, a more detailed comparison of the HRG model
with STAR data reveals that at high energies, deviations can
not be excluded \cite{karsch3,star2}.

All moments of net-proton fluctuations as well as the related cumulants
can be calculated once the underlying probability distribution is known.
Therefore, it  is interesting to confront the
net-proton distributions obtained for the HRG with those measured in heavy ion collisions. 
Since the probability
distributions contain information on all cumulants, such a
comparison may
provide useful insights into the origin of possible deviations from the HRG 
baseline.
An analysis of this kind may also provide additional  information on the relation between
chemical freeze-out and the QCD cross-over transition or even on the
existence of a critical end point in the QCD phase diagram.

In the following we calculate the net-proton  probability distributions in
the HRG model.
We show that in this model the net-proton distribution can be
expressed solely in terms of the measurable yields of protons and anti-protons.
The HRG model results are compared with data taken by the STAR Collaboration
in Au-Au collisions at RHIC
at  $\sqrt{ s_{NN}}=$  200 GeV. For central collisions
at this energy, the HRG model yields a distribution which  is
broader than the one seen experimentally. We discuss the possible origin of
the deviations and point out that they are in qualitative agreement
with the differences between HRG model calculations and (lattice) QCD
results found in the vicinity of the QCD phase boundary and above.

 \setcounter{equation}{0}
\section{Probability distribution of conserved charges }
%
Consider a sub-volume $V$ of a thermodynamic system described by the grand 
canonical ensemble consisting of charged particles $q$ and anti-particles $\bar{q}$ 
at a given temperature $T$ and chemical potential $\mu$. The latter is related to the conserved 
net charge $N=N_q-N_{\bar q}$. 
The probability distribution $P(N)$ for finding 
a net charge number $N$ in the volume $V$ is given in terms of 
the canonical $Z(T,V,N)$ and grand canonical
$\mathcal{Z}(T,V,\mu)$ partition functions \cite{hwa,koch,turko},
\begin{equation}
P(N) =  Z(T,V,N) e^{\hat{\mu} N - V\beta  p(T, \mu) }\; .
\label{eq5}
\end{equation}
Here we have introduced the thermodynamic pressure,
$\ln \mathcal{Z}=V\beta \, p(T,\mu)$, as well as  the shorthand notation 
$\hat\mu=\beta \mu$ and $\beta=1/T$.

The canonical partition function $Z(T,V,N)$ can be obtained from
the thermodynamic pressure through the discrete Fourier transform
\begin{equation}
Z(T,V,N)={1\over {2\pi}}\int_0^{2\pi} d\phi e^{-i\phi N} e^{\beta V p(T,i\phi/\beta) } \; ,
\label{eq6}
\end{equation}
where the  chemical potential  was Wick rotated
by the substitution $\hat\mu\to i\phi$.
Eqs.~(\ref{eq5}) and (\ref{eq6}) define the probability distribution of
a conserved charge in a sub-volume $V$  of a thermal  system described by the thermodynamic
pressure $p(T,\mu)$.


In the following we focus on fluctuations of the net-baryon number
and consider the corresponding probability distribution
in a strongly interacting medium.
We model the thermodynamics of this system using the HRG partition function, 
which contains all relevant degrees of
freedom in the hadronic phase  and implicitly includes the interactions
responsible for resonance formation.
In this model the thermodynamic pressure is a sum of meson and baryon contributions. 
This implies that only the baryonic
pressure $p_B$ contributes to  the probability distribution of the net-baryon number.
In the HRG model  $p_B$ consists of contributions from all baryons and
baryonic resonances \cite{hwa}. In the Boltzmann approximation,
$\beta V p_B(T,\mu)=b(T,V,\mu)+\bar b(T,V,\mu)$, where
$b$ and $\bar b$  is the mean number of
baryons and  anti-baryons  respectively,
\begin{eqnarray}
b(T,V,\mu)=\frac{VT}{2\pi^2} \sum_{i\in {\rm baryons}}
\hspace*{-0.1cm}
g_i m_i^2 K_2(\beta m_i)
 e^{\hat\mu+\beta\vec q_i\vec\mu }  \; ,
\label{eq7}
\end{eqnarray}
Here $\vec q_i=(S_i,Q_i)$ is a two-component vector composed of the strangeness and electric charge  carried by particle $i$, $\vec \mu_q=(\mu_S,\mu_Q)$ the  corresponding chemical potential vector, $g_i$ the spin-isospin degeneracy factor and $K_{2}$ is a modified Bessel function.
The mean number of
anti-baryons $\bar{b}$ is obtained by the
substitution  $\mu\to -\mu$ for all relevant chemical potentials.

In the hadron resonance gas model, the canonical partition function $Z(T,V,N)$  is
computed directly from  Eq.~(\ref{eq6}), using the thermodynamic
pressure discussed above~\cite{hwa}.
The resulting probability distribution of net-baryon number can be 
expressed solely in terms of the mean number of baryons and
anti-baryons,
\begin{eqnarray}
P(N) = \left( \frac{b}{{\bar b}}\right)^{N/ 2}
{{I_N(2\sqrt {b\bar b}~)}
 ~{\exp[-(b +{\bar b})]}} \; ,
\label{eq14}
\end{eqnarray}
where $I_N(x)$ is a modified Bessel function.

We note that the above arguments remain valid also for subsystems
which are limited not only in position space, but more generally in phase space. In particular, 
the introduction of cuts in momentum space leave Eqs.~(\ref{eq5}), (\ref{eq6}) 
and (\ref{eq14}) unchanged after an appropriate redefinition of the partition functions and densities.
Moreover, the restriction to one particle species, e.g. protons with proper account for resonance decays, 
is easily accommodated. 

\section{Probability distribution of the net-proton number}

In this section we confront the HRG model results for the probability distribution of
the net-proton number with data
of the STAR Collaboration \cite{star1,star2}.
The data is obtained at mid-rapidity
in a restricted range of transverse momentum,
$0.4~{\rm GeV} \le p_T \le 0.8~{\rm GeV}$.
The probability distribution for protons is readily obtained  from Eq.~(\ref{eq14}), by replacing 
the  mean number of baryons $b$  and anti-baryons $\bar b$
by that of protons
$\langle N_p\rangle $ and anti-protons $\langle N_{\overline p}\rangle$,
respectively.

Using Eq.~(\ref{eq14})  we can compute the net-proton
distribution provided we have access to the mean values $\langle N_p\rangle $
and  $\langle N_{\bar p}\rangle$ measured in the same kinematic window.
For Au-Au collisions at $\sqrt{s_{NN}}=$ 200 GeV, STAR  data on the $p_t$-distribution of protons and anti-protons 
is available at several centralities \cite{star3}.
By integrating the  $p_t$-spectra of anti-protons
in the $p_t$-window
where the net-proton number  was obtained,
we find:
$\langle N_{\bar p}\rangle =5.233(95)$,
$\langle N_{\bar p}\rangle=1.838(7)$ and
$\langle N_{\bar p}\rangle=2.844(98)$
for  (0-5$\%$)-central,
(30-40)$\%$-mid-central and (70-80)$\%$-peripheral collisions, respectively.
Since the proton data were measured in a slightly larger $p_t$ window, we
avoid systematic
errors that may arise from an extrapolation by computing
$\langle N_p\rangle$ from the net-proton yields
($M=\langle N_p\rangle - \langle N_{\bar p}\rangle $),
with 
$M\simeq 1.715$, $M\simeq 0.597$ and  $M\simeq 0.08$  for central,
mid-central and peripheral collisions~\cite{star1}.

In Fig.~\ref{fig1} (left) we compare
the STAR data on the net-proton multiplicity distribution in Au+Au
collisions at $\sqrt{s_{NN}}=$200 GeV \cite{star1,star2}
with the probability distributions obtained
in the HRG model (Eq.~({\ref{eq14})), using the experimental data
on $M$ and $\langle N_{\bar{p}}\rangle $ as input.
The data correspond to several centrality bins in the rapidity window
$| y |< 0.5$.
The distribution in Eq.~({\ref{eq14})  is normalized to unity.
In order to confront the HRG model with data on an absolute scale, we
adjust the normalization to that of the experimental
data in each centrality bin.

Fig.~\ref{fig1}~(left) shows  that in
peripheral collisions  at $\sqrt{s_{NN}}=$200 GeV the HRG model results agree
well with the data.
However, with increasing centrality deviations develop; in central collisions the
hadron resonance gas yields a distribution, which is
broader than the experimental one.
Below we argue that such deviations
are expected if the freeze-out conditions probed by fluctuations are located close
to the QCD cross-over temperature.  Consequently, the deviations observed in
Fig.~\ref{fig1} (left) and the dependence on centrality, could be
an indication for critical behavior.

We have also calculated
the number of protons and anti-protons  
with chemical freeze-out parameters adjusted to the multiplicities  
measured in Au-Au collisions at $\sqrt{ s_{NN}}=$200 GeV. The centrality dependence
of the freeze-out parameters were determined in Ref.~{\cite{star3}.
In this approach an additional input parameter is required, the
effective volume ($V$), which we fix by requiring that
the measured net-proton number $M$ at chemical freeze-out
is reproduced by the HRG model. As shown in Fig.~\ref{fig1}~(left)
the probability distributions obtained in this way are
consistent with  those computed directly from the measured  yields of
protons and anti-protons using Eq.~(\ref{eq14}).
The consistency of the two approaches strengthens our conclusion that 
the HRG model does not describe the net-proton probability distribution $P(N)$
for central Au-Au collision at $\sqrt{ s_{NN}}=$ 200 GeV.

Recently the STAR Collaboration also presented preliminary
data on the net-proton distribution in Au-Au collisions at
$\sqrt{ s_{NN}}=$ 39 GeV for various centralities \cite{star2}.
It would be interesting to  compare this data with the
HRG model. In this case, however, the corresponding data on
$p_t$-distributions, which would allow one to determine
$\langle N_p\rangle $  and $\langle N_{\bar p}\rangle $ directly from
the experiment is not  available. Thus, here we can only follow the second approach, i.e.,
employ the measured net-proton number $M$ and the 
chemical freeze-out parameters from Ref.~\cite{cleymans} to determine
$\langle N_p\rangle $  and $\langle N_{\bar p}\rangle $, which are then used as
input for the calculation of $P(N)$ based on Eq.~(\ref{eq14}).

A first comparison of the HRG model calculation
with the probability distribution obtained
for central Au-Au collisions at $\sqrt{ s_{NN}}=$ 39~GeV
indicates that at this energy
the shape and magnitude of the measured net-proton
distribution, and consequently  also the first four  measured moments,
are described well by the HRG model.
This suggests that in central Au-Au collisions at $\sqrt{ s_{NN}}=$ 39 GeV
the fluctuations as well as the particle yields are
characterized by the thermodynamic freeze-out conditions
corresponding to the statistical operator of the hadron resonance gas model.
Clearly, this result needs to be confirmed
by the final STAR data at this lower energy. If correct, this would
suggest that the deviations from the HRG results grow with
increasing energy. This feature can also be tested at the LHC.

The observed deviations in the probability distribution are also manifested in
differences between calculated and measured cumulants of the net-proton
fluctuations. In
Fig.~\ref{fig1} (right)  we show the  mean, variance,  skewness and kurtosis
obtained from the probability distributions  in
Fig.~{\ref{fig1}}~(left).
The HRG model, with experimental input for proton
and anti-proton yields a slightly better description of all four moments. However, a 
good overall description is obtained only for peripheral collisions.

The fact, that the HRG model yields a distribution which is broader than the experimental one implies
that deviations arise already on the level of the second order cumulant
(variance), which has the smallest experimental error.
This is expected if the particle freeze-out occurs near the
QCD cross-over transition.
In the cross-over region the baryon number
susceptibility ($\chi_2^B$), i.e., the second order cumulant
$\sigma^2 = VT^3\chi_2^B$ \cite{karsch3}, keeps rising steeply 
 with temperature
in the HRG model, while in QCD calculations it bends over and eventually
approaches a finite value at high temperatures. In the Gaussian approximation
(the leading order cumulant expansion) to the probability  distribution,
$P(N)\sim \exp [-N^2/(2 \sigma^2)]$, this implies that the distribution
in QCD is narrower than in the HRG model.

A possible interpretation of this effect may be related to the proximity of the freeze-out and cross-over regions probed at the highest
beam energy. Lattice calculations suggest that at larger $\mu$, 
the freeze-out curve the cross-over transition separate~\cite{karsch2,Endrodi}. Hence, one expects that the fluctuations reflect the critical dynamics at the cross-over transition only at the highest beam energies.    

 \section{Conclusions}
We have analyzed properties of  the net-proton  probability distributions in
heavy ion collisions within the hadron resonance gas model. In this model these  distributions can be expressed solely in terms of
the mean numbers of protons and anti-protons in a thermal system.
This provides a direct and unambiguous way to compare experimental data with model
predictions.

We have shown that the HRG model  describes the
data obtained by STAR in Au-Au collisions at
 $\sqrt{ s_{NN}}=$~200~GeV
for peripheral events
but differ for central ones.
Since the probability distributions are computed directly from
the measured proton and anti-proton yields, this deviation is not due to uncertainties in model
parameters. We suggest that this effect could be due to
the proximity of the freeze-out and cross-over regions at the highest
beam energy. In order to substantiate
this interpretation, data on the
net-proton distribution  at lower RHIC energies and
at LHC energies are needed.\\[5mm]

We acknowledge stimulating  discussions with
X. Luo, T. Nayak,
J.  Stachel, L. Turko,  Nu Xu and members of ALICE Collaboration.
K.R. received
partial support of the
Polish
Ministry of National Education (MEN).  The work of F.K. was
supported in part by contract DE-AC02-98CH10886 with the U.S. Department of
Energy.


\end{document}